\documentclass[prd,twocolumn,showpacs,floatfix,amsmath,amssymb,nofootinbib]{revtex4}

\usepackage{amsmath,amssymb,mathrsfs,bm}
\usepackage{cases}

\usepackage{graphicx}
\usepackage{overpic}
\usepackage{booktabs}
\usepackage{multirow}
\usepackage{array}

\usepackage{color}
\usepackage{indentfirst}

\usepackage[colorlinks,
citecolor=blue,
linkcolor=red,
urlcolor=blue]{hyperref}
\usepackage{cleveref}

\newcolumntype{C}[1]{>{\centering\arraybackslash}p{#1}}

\begin{document}
\title{Dynamical study of hidden-strange pentaquarks as analogs of the hidden-charm states}

\author{Xuejie Liu$^1$}\email[E-mail: ]{1830592517@qq.com}
\author{Yue Tan$^{2}$}\email[E-mail:]{tanyue@ycit.edu.cn}
\author{Yuheng Wu$^{2}$}\email[E-mail:]{wuyuheng@ycit.edu.cn}
\author{Dianyong Chen$^{3,5}$\footnote{Corresponding author}}\email[E-mail:]{chendy@seu.edu.cn}
\author{Hongxia Huang$^4$}\email[E-mail:]{hxhuang@njnu.edu.cn}
\author{Jialun Ping$^4$}\email[E-mail: ]{jlping@njnu.edu.cn}
\affiliation{$^1$School of Physics, Henan Normal University, Xinxiang 453007, P. R. China}
\affiliation{$^2$School of Mathematics and Physics, Yancheng Institute of Technology, Yancheng, 224051,  P. R. China}
\affiliation{$^3$Lanzhou Center for Theoretical Physics, Lanzhou University, Lanzhou 730000, P. R. China}
\affiliation{$^4$Department of Physics, Nanjing Normal University, Nanjing 210023, P. R. China}
\affiliation{$^5$School of Physics, Southeast University, Nanjing 210094, P. R. China}

\begin{abstract}
Motivated by the recent BESIII experiment~\cite{BESIII:2024muk} searching for hidden-strange exotic hadrons, we perform a systematic theoretical study of the hidden-strange pentaquark system within the framework of the quark delocalization color screening model (QDCSM) and the resonating group method (RGM). Our results demonstrate that the channel coupling effect plays a decisive role in the formation of bound and resonance states. It not only significantly enhances the short-range attraction but also induces essential attractive contributions from pion exchange. We predict three bound states with masses of $1759$ MeV, $2000$ MeV, and $2407$ MeV. Furthermore, we report the existence of a hidden-strange pentaquark resonance state, $\Xi K^{\ast}$, with quantum numbers $I(J^{P})=0(1/2^{-})$. This resonance is identified in the $S$-wave scattering phase shifts of the $\Lambda\eta_{s}$ and $\Lambda \phi$ open channels, with a predicted mass in the range of $2204\text{--}2208$ MeV. By accounting for both the scattering width from channel coupling and the intrinsic decay width of the constituent $K^{\ast}$, the total decay width is estimated to be $55\text{--}63$ MeV. These theoretical predictions provide important guidance for future experimental searches for such exotic states at facilities like BESIII.
\end{abstract}

\maketitle

\section{\label{sec:introduction}Introduction}
The exploration of multiquark exotic hadron states that deviate from the conventional hadron classification (i.e., baryons composed of three quarks and mesons made of a quark-antiquark pair)-has consistently been one of the most significant and active research topics in the hadron physics.  Experimental searches for hidden-strange pentaquark states began as early as the 1990s. The SPHINX Collaboration, analyzing proton diffraction reactions $p+N(C)\rightarrow X(\Sigma^{0} K^{+})\rightarrow N(C)$, reported an X(2000) structure and proposed it as a $uuds\bar{s}$ state candidate~\cite{Naftulin:1994zn,SPHINX:1995wtz,SPHINX:1994yxk,SPHINX:1999rcq}. Although this observation had limited statistical significance, it stimulated some theoretical studies for the existence of the hidden-strange pentaquarks and triggered in-depth discussion on various theoretical models. However, subsequent experiments, such as the Belle Collaboration's search in  $\Lambda_c^+ \rightarrow \phi p \pi^0$ decays, found no evidence of a $P_s$ intermediate state decaying to $\phi p$~\cite{Belle:2017tfw}. Consequently, whether hidden-strange pentaquark states exist remained a controversial and unresolved issue for a long time. 

The landscape of exotic hadron physics was revolutionized in 2015. The LHCb Collaboration first reported the observation of two exotic states with hidden-charm, $P_{c}(4380)$ and $P_{c}(4450)$, through an analysis of the $\Lambda_{b}\rightarrow J/\psi p K$decay~\cite{LHCb:2015yax,LHCb:2016lve}. Later, utilizing a larger data sample in the same process in 2019,  LHCb further discovered three narrow structures,  $P_c(4312)$, $P_c(4440)$, and $P_c(4457)$, with $P_c(4440)$ and $P_c(4457)$ interpreted as the split of $P_c(4450)$~\cite{LHCb:2019kea}. 
Subsequently, the hidden-charm resonance $P_{c}(4337)$ was also confirmed in the mass distribution of $J/\psi p$ and $J/\psi \bar{p}$ in the $B_{s}\rightarrow J/\psi p \bar{p}$ decay process~\cite{LHCb:2021chn}. Further searches for similar hidden-charm pentaquark candidates in other decay process, such as $\Lambda_{b}\rightarrow \eta_{c} p K$, yielded no positive signals~\cite{LHCb:2020kkc}. Additionally, two hidden-charm strange exotic states $P_{cs}(4459)$ and $P_{cs}(4338)$ were also detected in the $ J/\psi \Lambda$  mass distribution in the $\Xi_{b}\rightarrow J/\psi \Lambda K$~\cite{LHCb:2020jpq} and $B^{-}\rightarrow J/\psi \Lambda p$ decay~\cite{LHCb:2022ogu}. The uncertain property and structure of these discovered $P_{c}$  and $P_{cs}$ states-whose minimal quark content ($uud c\bar{c}$/$uds c\bar{c}$) classifies them as pentaquarks-have sparked significant interest in the theoretical community. Among the various proposed theoretical interpretations, the baryon-meson molecular picture is widely favored~\cite{Chen:2016heh,Chen:2016otp,Azizi:2016dhy,Azizi:2018bdv,Azizi:2020ogm,He:2019ify,Zhu:2021lhd,Liu:2019tjn,Xiao:2019mvs,Lu:2016nnt,Hu:2021nvs,Meng:2019dba,Chen:2015loa,Chen:2015moa,He:2015cea,Meissner:2015mza,Roca:2015dva,Chen:2020opr,Yan:2021nio,Wu:2021caw,Chen:2020uif,Phumphan:2021tta,Du:2021fmf,Lu:2021irg,Gao:2021hmv,Yalikun:2021dpk}, primarily because these observed $P_{c}$  and $P_{cs}$ states are located near the corresponding hadron-pair thresholds. Nevertheless, alternative explanations, including hadron-charmonia~\cite{Eides:2019tgv}, compact pentaquark states~\cite{Ali:2019npk,Wang:2019got,Cheng:2019obk,Weng:2019ynv,Zhu:2019iwm,Pimikov:2019dyr,Ruangyoo:2021aoi}, virtual states~\cite{Fernandez-Ramirez:2019koa}, triangle singularities~\cite{Nakamura:2021qvy}, and cusp effects~\cite{Burns:2022uiv}, remain plausible. The discoveries of $P_{s}$ and $P_{cs}$ states established the existence of genuine pentaquark states and unveiled a hidden-charm pentaquark family with different strangeness numbers.

\begin{figure}[t]
    \centering
   \includegraphics[width=0.95\columnwidth]{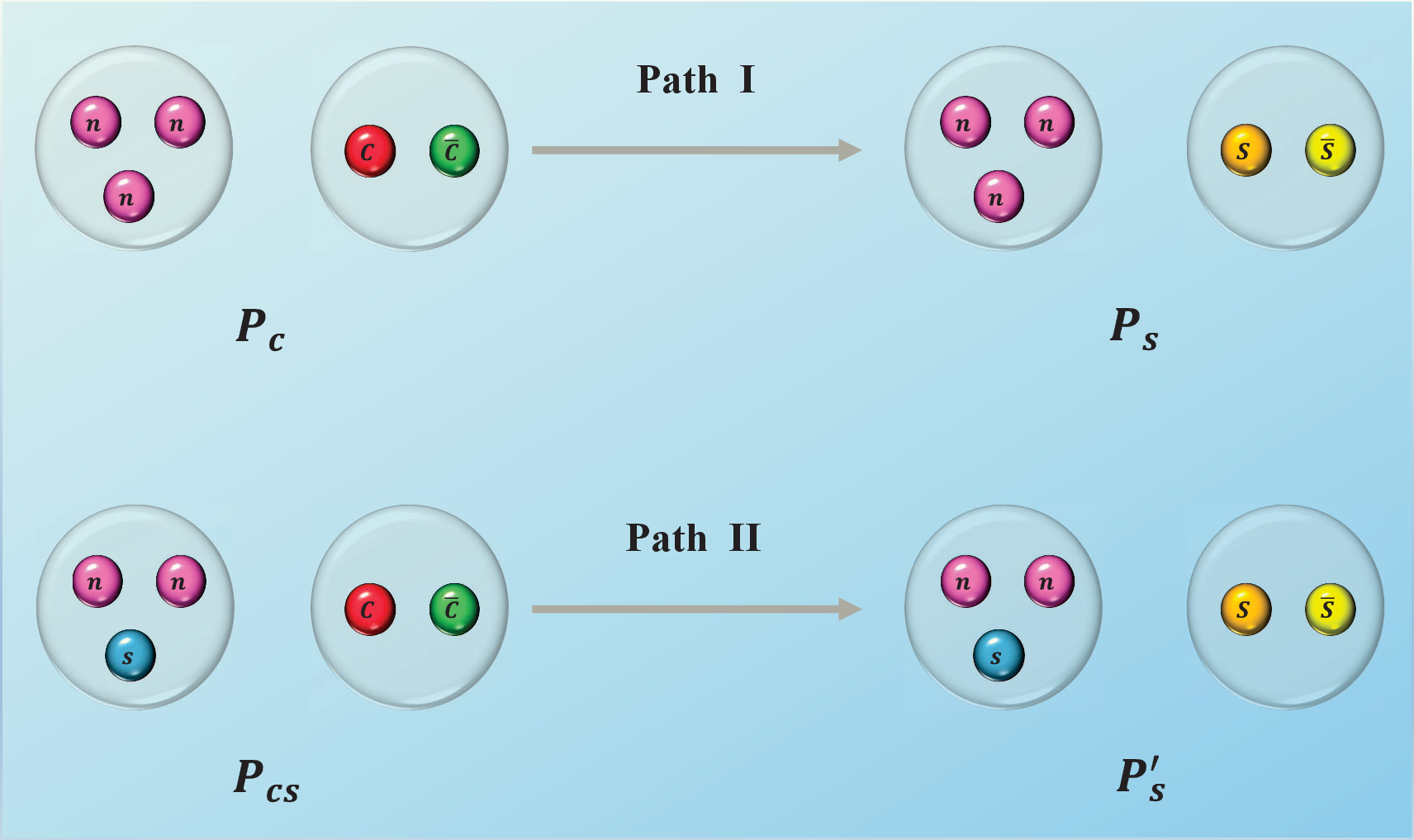}
    \caption{Two parallel paths analogous to the hidden-charm sector for investigating hidden-strange pentaquark states, where $"n"$ denotes a $u$ or $d$ quark. }
    \label{Ps-ps}
\end{figure}

 Significant breakthroughs in the hidden-charm sector provide a compelling impetus to re-examining the unresolved hidden-strange pentaquark question. Guided by flavor symmetry, the rich spectrum of $P_c$ and $P_{cs}$ states suggests the existence of analogous hidden-strange multiplets. As illustrated in Fig.~\ref{Ps-ps}, this analogy opens two parallel paths for investigation.  The first analogy stems from the correspondence between $P_c$ and $P_s$ states: the experimental confirmation of $P_{c}$ states has strongly motivated the search for their hidden-strange counterparts, $P_{s}$ states, which were the original focus of the SPHINX experiment. In this context, many theoretical studies have followed. For instance, Ref.~\cite{Williams:2003tj} predicted four $P_{s}$ candidates within the non-relativistic quark model. In Ref.~\cite{Gao:2000az}, a $N\phi$ bound state was proposed due to the QCD van der Waals attractive potential. In Ref.~\cite{Lebed:2015dca}, the results of the $\Lambda_{c}^{+}\rightarrow \pi^{0} \phi p$ decay process based on the triangle singularity mechanism, also suggested the potential existence of $P_{s}$ resonance states. Furthermore, investigations within the chiral SU(3) quark model~\cite{Huang:2005gw,Liu:2018nse}, lattice QCD~\cite{Beane:2014sda}, chiral soliton model~\cite{Kopeliovich:2015vqa}, QCD sum rule~\cite{Ben:2025wqn} have also supported the existence of $P_{s}$ states. Notably, our previous works using the QDCSM framework~\cite{Gao:2017hya,Huang:2018ehi} also predicted several $P_s$ bound states. These QDCSM predictions are qualitatively consistent with the recent results from QCD sum rules~\cite{Ben:2025wqn} and other effective models, demonstrating that the QDCSM is a reliable framework for describing multiquark systems with strange content.

The second exploratory path, which is the focus of this work,  draws an analogy between $P_{cs}$ states and potential $P_{s}^{'}$ states.  The experimental discovery of the $P_{cs}$ states has naturally inspired searches for its hidden-strange analogue, $P_{s}^{'}(udss\bar{s})$.  Recently, the BESIII Collaboration reported the first investigation of the $\psi(3686)\rightarrow \Sigma^{0}\bar{\Sigma}^{0}\phi$ decay with a statistical significance of $7.6\sigma$, simultaneously examining possible intermediate resonances in the $\Sigma^{0}\phi$ ($\bar{\Sigma}^{0}\phi$) invariant-mass spectra~\cite{BESIII:2024muk}. Although no statistically significant signal was observed, the results do not rule out the possible existence of such hidden-strange resonance states. This series of important experimental advances, from the discovery of $P_{cs}$ to the exploration of hidden-strange structures, highlights the urgent need for a unified theoretical framework to systematically predict the hidden-strange pentaquark signals in strange sectors. Compared to the extensive literature on the $P_s$ system, dynamical studies of the $P_{s}^{'}$ system are remarkably scarce.
 Notably, a recent QCD sum rule study suggested a possible $P_{s}'$ state with $J^{P}=3/2^{-}$~\cite{Ben:2025wqn}, but a comprehensive dynamical calculation involving coupled channels is still lacking.

 Inspired by the above analysis, we perform a comprehensive dynamical study of the hidden-strange pentaquark with strangeness system within the framework of the quark delocalization color screening model. The possible physical channels with different quantum numbers are considered.  To explore the interaction of hadron-hadron pair, the effective potential of each channel is estimated. Then, to confirm whether the physical channel with effective attractive potential yields a bound-state solution through the resonating group method in the dynamical estimation, the corresponding single channel and channel coupling estimations are applied to this system. Meanwhile, with the consideration of quantum number conservation and phase space constraints, the hadronic decay process of the hidden-strange pentaquark are investigated to identify possible resonance states and estimate their decay widths.  These predictions aim to  provide a complete theoretical picture for future experimental searches.

This work is organized as follows. After the introduction, the detail of QDCSM is given in Sec.~\ref{mod}. A comprehensive numerical analysis of the hidden-strange pentaquark with strangeness is carried out, including the effective potential, possible bound states, and resonance states in Sec.~\ref{dis}, and a brief summary of this work is given in Sec.~\ref{sum}.

\section{THE QUARK DELOCALIZATION COLOR SCREENING MODEL  \label{mod}}

The present work investigates hidden-strange pentaquark configurations with quark content $nnss\bar{s}$ (where $n$ represents $u$ or $d$ quark) within the quark-delocalization color-screening model (QDCSM), solved by the resonating-group method (RGM). The QDCSM extends the nonrelativistic quark-cluster model to multiquark systems by incorporating orbital delocalization and an effective color-screening between clusters, which effectively encodes hidden-color-channel dynamics.  Detailed descriptions of the QDCSM can be found in Refs.~\cite{Wang:1992wi, Chen:2007qn, Chen:2011zzb, Wu:1996fm, Huang:2011kf}.

In the QDCSM, the general form of the Hamiltonian for the pentaquark system is,
\begin{equation}
H = \sum_{i=1}^{5} \left(m_i+\frac{\boldsymbol{p}_i^2}{2m_i}\right)-T_{\mathrm{CM}}+\sum_{j>i=1}^5V(r_{ij}),\\
\end{equation}
where the center-of-mass kinetic energy, $T_{\mathrm{CM}}$, is subtracted without losing generality since we mainly focus on the internal relative motions of the multiquark system.  
The two-body interaction
\begin{equation}
V(r_{ij})=V_{\rm CON}(r_{ij})+V_{\rm OGE}(r_{ij})+V_{\chi}(r_{ij})
\end{equation}
contains the color-confining potential, $V_{\mathrm{CON}}$, one-gluon exchange potential, $V_{\mathrm{OGE}}$, and Goldstone-boson exchange potential, $V_{\chi}$, respectively.  Note that these potentials include the central, spin-spin, spin-orbit, and tensor contributions, respectively. However, as the present analysis focuses on $S$-wave states, only the first two kinds of potentials will be considered attending the goal of the present calculation and for clarity in our discussion. 

 In particular, the one-gluon-exchange potential, $V_{\mathrm{OGE}}(r_{ij})$, is written as
\begin{eqnarray}
V_{\mathrm{OGE}}(r_{ij}) &=& \frac{1}{4}\alpha_{s} \boldsymbol{\lambda}^{c}_i \cdot\boldsymbol{\lambda}^{c}_j \nonumber\\
&&\times\left[\frac{1}{r_{ij}}-\frac{\pi}{2}\delta(\boldsymbol{r}_{ij})\left(\frac{1}{m^2_i}+\frac{1}{m^2_j}
+\frac{4\boldsymbol{\sigma}_i\cdot\boldsymbol{\sigma}_j}{3m_im_j}\right)\right],\ \
\end{eqnarray}
where $m_{i}$ is the quark mass, $\boldsymbol{\sigma}$ and $\boldsymbol{\lambda^{c}}$ are the Pauli matrices and SU(3) color matrices, respectively. The QCD-inspired effective scale-dependent strong coupling constant, $\alpha_{s}$, offers a consistent description of mesons and baryons from the light to the heavy quark sectors


In the QDCSM, the confining interaction $V_{\mathrm{CON}}(r_{ij})$ can be expressed as
\begin{equation}
 V_{\mathrm{CON}}(r_{ij}) =  -a_{c}\boldsymbol{\lambda^{c}_{i}\cdot\lambda^{c}_{j}}\Big[f(r_{ij})+V_{0_{ij}}\Big] \ ,
\end{equation}
where $a_{c}$ represents the strength of the confinement potential and $V_{0_{ij}}$ refers to the zero-point potential. Furthermore, in the quark delocalization color screening model, the quarks in the considered pentaquark state involving quark components of $nnss\bar{s}$ are first divided into two clusters, which are baryon cluster composed of three quarks, and meson cluster composed of one quark and one antiquark. Consequently, the five-body problem is simplified into a two-body problem. The delocalization function $f(r_{ij})$ is defined as:
\begin{equation}
 f(r_{ij}) =  \left\{ \begin{array}{ll}r_{ij}^2 & \quad \mbox{if }i,j\mbox{ occur in the same cluster}, \\
\frac{1 - e^{-\mu_{ij} r_{ij}^2} }{\mu_{ij}} & \quad \mbox{if }i,j\mbox{ occur in different cluster},
\end{array} \right.
\label{Eq:fr}
\end{equation}
where the color screening parameter $\mu_{ij}$ is determined by fitting the deuteron properties, nucleon-nucleon, and nucleon-hyperon scattering phase shifts~\cite{Chen:2011zzb, Wang:1998nk}, with $\mu_{nn}= 0.45\ \mathrm{fm}^{-2}$, $\mu_{ns}= 0.19\ \mathrm{fm}^{-2}$
and $\mu_{ss}= 0.08\ \mathrm{fm}^{-2}$, satisfying the relation $\mu_{ns}^{2}=\mu_{nn}\mu_{ss}$, where $n$ represents $u$ or $d$ quark.  It should be noted that $\mu_{ij}$ are phenomenological model parameters, their values are determined by reproducing the relevant mass spectra and phase shifts of the scattering processes. In Ref.~\cite{Chen:2011zzb}, it was found that the relation $\mu_{qs}^2=\mu_{qq} \mu_{ss}$ allows for a good reproduction of the ground state baryon masses in the light quark sector.  Subsequently, such relations have been successfully applied to investigate states containing heavy quarks~\cite{Huang:2013rla, Huang:2015uda, Liu:2022vyy}.

The Goldstone-boson exchange interactions between light quarks appear because of the dynamical breaking of chiral symmetry. The following $\pi$, $K$, and $\eta$ exchange terms work between the chiral quark-(anti)quark pair, which read,
\begin{eqnarray}
V_{\chi}(r_{ij}) & =&  v^{\pi}_{ij}(r_{ij})\sum_{a=1}^{3}\lambda_{i}^{a}\lambda_{j}^{a}+v^{K}_{ij}(r_{ij})\sum_{a=4}^{7}\lambda_{i}^{a}\lambda_{j}^{a}+v^{\eta}_{ij}(r_{ij})\nonumber\\
&&\left[\left(\lambda _{i}^{8}\cdot
\lambda _{j}^{8}\right)\cos\theta_P-\left(\lambda _{i}^{0}\cdot
\lambda_{j}^{0}\right) \sin\theta_P\right], \label{sala-Vchi1}
\end{eqnarray}
with
\begin{eqnarray}
  v^{GB}_{ij} &=&  {\frac{g_{ch}^{2}}{{4\pi}}}{\frac{m_{GB}^{2}}{{\
12m_{i}m_{j}}}}{\frac{\Lambda _{GB}^{2}}{{\Lambda _{GB}^{2}-m_{GB}^{2}}}}
m_{GB}     \nonumber    \\
&&\times\left\{(\boldsymbol{\sigma}_{i}\cdot\boldsymbol{\sigma}_{j})
\left[ Y(m_{GB}\,r_{ij})-{\frac{\Lambda_{GB}^{3}}{m_{GB}^{3}}}
Y(\Lambda _{GB}\,r_{ij})\right] \right\},
\end{eqnarray}
the subscript GB in $m_{GB}$ and $\Lambda_{GB}$ represents the mesons $\pi$, $ K$, and  $\eta$, while $m_{GB}$ denotes their respective masses, as derived from the experimental value~\cite{ParticleDataGroup:2018ovx} and $\Lambda_{GB}$ serves as the cutoff parameter.  $Y(x)=e^{-x}/x$ is  the standard Yukawa function. $\boldsymbol{\lambda^{a}}$ is the SU(3) flavor Gell-Mann matrix. By matching the pion exchange diagram of the $NN$ elastic scattering process at the quark level and the hadron level, one can relate the $\pi qq$ coupling with the one of $\pi NN$, which is~\cite{Vijande:2004he, Fernandez:1986zn},
\begin{equation}
\frac{g_{ch}^{2}}{4\pi}=\left(\frac{3}{5}\right)^{2} \frac{g_{\pi NN}^{2}}{4\pi} {\frac{m_{u,d}^{2}}{m_{N}^{2}}},
\end{equation}
which assumes that the flavor SU(3) is an exact symmetry, and only broken by the masses of the strange quark. As for the coupling $g_{\pi NN}$, it was determined by the $NN$ elastic scattering~\cite{Fernandez:1986zn}. All model parameters for the light and strange sectors follow Ref.~\cite{Huang:2018ehi} , where they were determined by reproducing the mass spectrum of the ground states mesons and baryons within QDCSM.  Guided by Ref.~\cite{Huang:2018ehi}, which predicts strange partner states of the $P_{c}$ family, the properties of hidden-strange pentaquark systems are examined in the next section.

Besides, in QDCSM, the quark delocalization is realized by specifying the single-particle orbital wave function as a linear combination of left and right Gaussian basis, the single--
particle orbital wave functions used in the ordinary quark cluster model read,
\begin{eqnarray}\label{wave0}
\psi_{\alpha}(\boldsymbol{s}_{i},\epsilon)&=&\left(\Phi_{\alpha}(\boldsymbol{s}_{i})
  +\epsilon\Phi_{\beta}(\boldsymbol{s}_{i})\right)/N(\epsilon), \nonumber \\
\psi_{\beta}(\boldsymbol{s}_{i},\epsilon)&=&\left(\Phi_{\beta}(\boldsymbol{s}_{i})
  +\epsilon\Phi_{\alpha}(\boldsymbol{s}_{i})\right)/N(\epsilon), \nonumber \\
N(\epsilon)&=& \sqrt{1+\epsilon^2+2\epsilon e^{-s^2_{i}/{4b^2}}},\nonumber \\
\Phi_{\alpha}(\boldsymbol{s}_{i})&=&\left(\frac{1}{\pi b^2}\right)^{\frac{3}{4}}
e^{-\frac{1}{2b^2}\left(\boldsymbol{r_\alpha}-\frac{2}{5}s_{i}\right)^2},\nonumber \\
\Phi_{\beta}(-\boldsymbol{s}_{i})&=&\left(\frac{1}{\pi b^2}\right)^{\frac{3}{4}}
e^{-\frac{1}{2b^2}\left(\boldsymbol{r_\beta}+\frac{3}{5}s_{i}\right)^2},
\end{eqnarray}
with $\boldsymbol{s}_{i}$, $i=(1,2,..., n)$, to be the generating coordinates, which are introduced to
expand the relative motion wave function~\cite{Wu:1998wu,Ping:1998si,Pang:2001xx}. The parameter $b$ indicates the size of the baryon and meson clusters, which is determined by fitting the radius of the baryon and meson by the variational method~\cite{Huang:2018rpb}. In addition, The mixing parameter $\epsilon(s_{i})$ is not an adjusted one but is determined variationally by the dynamics of the multi-quark system itself. This assumption allows the multi-quark system to choose its favourable configuration in the interacting process. It has been used to explain the crossover transition between the hadron phase\footnote{The phase shift of $NN$ interaction could be described with the formalisms with hadrons only. After including the pseudo-scalar, vector and scalar meson, especially the $\sigma$ meson, the $NN$ interaction has been well described. In Ref.~\cite{Chen:2007qn}, the authors concluded that the $\sigma$-meson exchange can be replaced by quark delocalization and color screening mechanism introduced by QDCSM by comparing the NN scattering and deuteron properties obtained by chiral quark model and QDCSM} and the quark-gluon plasma phase~\cite{Xu:2007oam, Chen:2007qn, Huang:2011kf}. 

\begin{table}[htb]
\begin{center}
\renewcommand\arraystretch{1.4}
\caption{\label{channels} The relevant $S$-wave channels for all possible states with different $J^P$ quantum numbers}
\begin{tabular}{C{0.8cm}C{1.cm}C{1.cm}C{1.cm}C{1.cm}C{1.cm}C{1.cm}C{1.0cm}}
\toprule[1pt]
$J^{P}$    &Isospin   &\multicolumn{6}{c}{Channels}  \\
\cmidrule(lr){3-8}
\midrule[1pt]
\multirow{2}{*}{$\frac{1}{2}^{-}$} &$I=0$  &$\Lambda \eta^{\prime}$    &$\Lambda \phi$     &$\Xi K$        &$\Xi K^{\ast}$     &$\Xi^{\ast}K^{\ast}$  &\\
                                                          &$I=1$  &$\Sigma\eta^{\prime}$       &$\Sigma \phi$        &$\Sigma^{\ast} \phi$  &$\Xi K$  &$\Xi K^{\ast}$  &$\Xi^{\ast} K^{\ast}$ \\
                                                       
\multirow{2}{*}{$\frac{3}{2}^{-}$} &$I=0$  &$\Lambda \phi$   &$\Xi K^{\ast}$  &$\Xi^{\ast} K$  &$\Xi^{\ast} K^{\ast}$  &  &\\
                                                          &$I=1$  &$\Sigma \phi$   &$\Sigma^{\ast} \eta^{\prime}$ &$\Sigma^{\ast} \phi$   &$\Xi K^{\ast}$   &$\Xi^{\ast} K$ &$\Xi^{\ast} K^{\ast}$ \\
                                                       
\multirow{2}{*}{$\frac{5}{2}^{-}$} &$I=0$  & $\Xi^{\ast} K^{\ast}$   &&&&& \\
                                                          &$I=1$   &$\Sigma^{\ast} \phi$  &$\Xi^{\ast} K^{\ast}$  &&&&\\                                        
    
\bottomrule[1pt]
\end{tabular}
\end{center}
\end{table}

\section{ Results and discussions\label{dis}}
\begin{figure*}[t]
    \centering
    \includegraphics[scale=1]{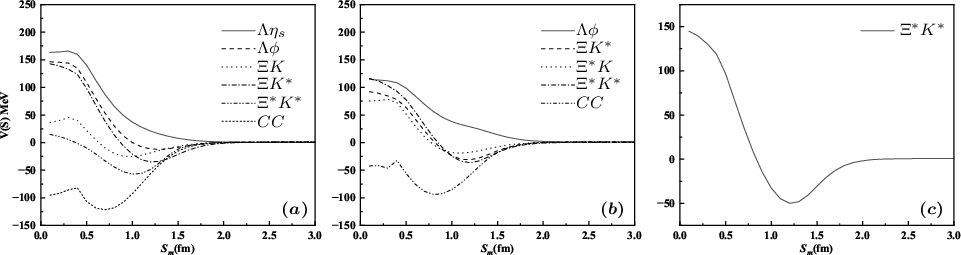}
    \caption{The effective potentials calculated within the QDCSM, as defined in Eq.~\ref{Eq:PotentialV}, for the $I=0$ hidden-strange pentaquark system. The panels correspond to the quantum numbers (a) $J^{P}=1/2^{-}$, (b) $J^{P}=3/2^{-}$, and (c) $J^{P}=5/2^{-}$, respectively. The  curve labeled $CC$ denotes the effective potential obtained from the coupled-channel calculation.}
    \label{I=0-effective}
\end{figure*}

\begin{figure*}[t]
     \centering
     \includegraphics[scale=1]{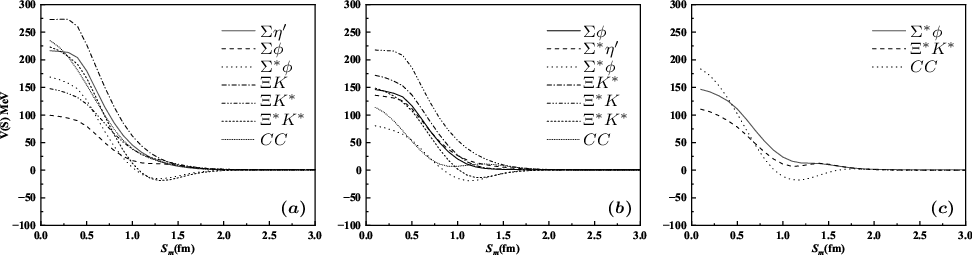}   
     \caption{The effective potentials for the $I=1$ hidden-strange pentaquark system; the notation is the same as in Fig.~\ref{I=0-effective}. The panels correspond to (a) $J^{P}=1/2^{-}$, (b) $J^{P}=3/2^{-}$, and (c) $J^{P}=5/2^{-}$. }
     \label{I=1-effective}
 \end{figure*}

In this work, a comprehensive study of low-lying hidden-strange pentaquarks with configuration $nnss\bar{s}$ is carried out within the quark delocalization color screening model(QDCSM). The analysis is restricted to $S$-wave channels with isospin $I=0$, $1$ and quantum numbers $J^{P}=\frac{1}{2}^{-}$, $\frac{3}{2}^{-}$, $\frac{5}{2}^{-}$. The corresponding physical channels are listed in Table~\ref{channels}. The primary objective of this work is to determine whether additional pentaquark states exist in this sector and to assess whether such states can be interpreted as hadronic molecules. To further characterize the properties of the pentaquark states, the effective potentials for each channel are derived and presented in Figs.~\ref{I=0-effective} and \ref{I=1-effective} as the first step, subsequently, based on the resonating group method (RGM) and generator coordinate method~\cite{Kamimura:1977okl,wheeler1937molecular,Hill:1952jb,Griffin:1957zza,Babenko:2003js}, we perform a dynamical bound-state calculation of both single channel and coupled-channel configurations to explore whether any other pentaquark states exist. The relevant results are listed in Tables~\ref{bound-1} and~\ref{bound-2}. Finally, according to the dynamic calculations, we investigate the effects of different channel coupling on the scattering process, with the main goal of finding and identifying resonance states. These results are presented in Table~\ref{decaywidth} and Fig.~\ref{phase-shifts}, respectively.

\subsection{The effective potential}
To search for bound states and resonance states within the hidden-strange pentaquark system, we first explore the effective potential of the hadron pair listed in Table~\ref{channels}, because the attractive potential is necessary for forming a bound state and resonance state. So the potential is defined as 
\begin{eqnarray}
    E(S_{m})&=&\frac{\langle\Psi_{5q}(S_m)|H|\Psi_{5q}(S_m)\rangle}{\langle\Psi_{5q}(S_m)|\Psi_{5q}(S_m)\rangle},\label{Eq:PotentialE}
\end{eqnarray}
where $S_m$ denotes the distance between two clusters and $\Psi_{5q}(S_m)$ represents the wave function of a certain given channel. The terms  $\langle\Psi_{5q}(S_m)|H|\Psi_{5q}(S_m)\rangle$ and $\langle\Psi_{5q}(S_m)|\Psi_{5q}(S_m)\rangle$ correspond to the Hamiltonian matrix and the overlap of the states, respectively. Thus, the effective potential between two colorless clusters is defined as
\begin{eqnarray}
    V(S_m)=E(S_m)-E(\infty), \label{Eq:PotentialV}
\end{eqnarray}
where $E(\infty)$ represents the energy at a sufficiently large distance between two clusters. The estimated potentials for $I=0$ and $I=1$ are presented in Fig.~\ref{I=0-effective} and Fig.~\ref{I=1-effective}, respectively.

Figure~\ref{I=0-effective} illustrates the effective potential for each channel in the $I=0$ system across different quantum numbers. For the $I(J^{P})=0(1/2^{-})$ system, as shown in Fig.~\ref{I=0-effective}(a), there are five physical channels: the $\Lambda\eta_{s}$, $\Lambda\phi$, $\Xi K$, $\Xi K^{\ast}$, and $\Xi^{\ast} K^{\ast}$, respectively.   One can see that the effective potentials for all physical channels are attractive except for the $\Lambda\eta_{s}$ channel, the effective potential of which exhibits  purely repulsive behavior. Therefore,  a bound state cannot be formed in the $\Lambda\eta_{s}$ channel within a single-channel estimation. Conversely, the remaining channels exhibit attractive properties, suggesting they are likely to support bound states in the dynamical calculation; these results are presented in the following section. When considering the channel coupling, the effective potential demonstrates strong attraction, indicating that  the channel coupling effect plays a crucial role in the potential formation of bound states in the $I(J^{P})=0(1/2^{-})$ system.

For the $I(J^{P})=0(3/2^{-})$ system,  the effective potentials of four physical channels: $\Lambda\phi$, $\Xi K^{\ast}$, $\Xi^{\ast}K$ and $\Xi^{\ast} K^{\ast}$ are shown in Fig.~\ref{I=0-effective}(b). They exhibit effective potential behavior similar to that of the $I(J^{P})=0(1/2^{-})$ system. The $\Lambda\phi$ shows purely repulsive behavior,  precluding the existence of a bound state in the single channel calculation. However, the attractive potentials of $\Xi K^{\ast}$, $\Xi^{\ast}K$ and $\Xi^{\ast} K^{\ast}$ indicate  that these channels are more likely to form bound states in the dynamical estimation. Additionally,  Fig.~\ref{I=0-effective}(b) presents the effective potential after channel coupling, where the observed strong attraction further supports the likelihood of bound state formation.

For the  $I(J^{P})=0(5/2^{-})$ system,  Fig.~\ref{I=0-effective}(c) shows only the $\Xi^{\ast}K^{\ast}$ channel due to the $S-$wave limitation.   The attractive effective potential of the $\Xi^{\ast}K^{\ast}$ channel can be observed at medium range,  implying that the $\Xi^{\ast}K^{\ast}$ is likely to form a bound state in the dynamical calculation. 

For the $I(J^{P})=1(1/2^{-})$ system, the effective potential behavior of the $\Sigma \eta_{s}$, $\Sigma\phi$, $\Sigma^{\ast}\phi$, $\Xi K$, $\Xi K^{\ast}$, and $\Xi^{\ast}K^{\ast}$ channels are presented in Fig.~\ref{I=1-effective}(a). The interaction in the  $\Xi^{\ast}K^{\ast}$ and $\Xi^{\ast}\phi$ channels exhibits very weak attraction at medium range, while a pronounced repulsive interaction dominates at close range. Given this scenario, these two channels may be unlikely to form stable bound states. For the $\Sigma \eta_{s}$, $\Sigma \phi$, $\Xi K$, and $\Xi K^{\ast}$ channels,  the effective potentials all exhibit repulsive characteristics, suggesting that the formation of bound states are unlikely within a single channel dynamical calculation.  Additionally, the repulsive potential can be seen after channel coupling, thus preventing the formation of bound states. 

A similar effective potential behavior can also be observed in the system with $I(J^{P})=1(3/2^{-})$. It is evident from Fig.~\ref{I=1-effective}(b) that channels $\Sigma^{\ast}\phi$ and $\Xi^{\ast} K^{\ast}$ exhibit weak attraction at the medium range (around 1.2 fm), but a strong repulsive interaction dominates at close range. This suggests that the formation of a bound state in these two channels is improbable. The remaining channels, specifically $\Sigma \phi$, $\Sigma^{\ast}\eta_{s}$, $\Xi K^{\ast}$, and $\Xi^{\ast} K$, consistently show strong repulsive behavior across both close and medium ranges, implying that these channels are also unlikely to form bound states. Therefore, based on the analysis of the effective potentials for the $I(J^{P})=1(3/2^{-})$ system, it is highly probable that no bound states emerge in the $I(J^{P})=1(3/2^{-})$ system.

Finally, for the  $I(J^{P})=1(5/2^{-})$ system, Fig.~\ref{I=1-effective}(c) reveals the presence of only channels $\Sigma^{\ast} \phi$ and $\Xi^{\ast} K^{\ast}$. Since the effective potentials of both channels  are repulsive, it is difficult for this system to form a bound state. Furthermore, while effective potential resulting from channel coupling exhibits weak attraction at the medium range, this attraction is insufficient to overcome the strong repulsive potential at close range. Consequently, no bound states are expected in this case.

  
\subsection{The dynamical calculation}

\begin{table*}[htb]
    \begin{center}
        \renewcommand{\arraystretch}{1.5}
        \caption{\label{bound-1}  Calculated masses and binding energies for the $I=0$ hidden-strange pentaquark system (in MeV). $E_{sc}$ and $E_{cc}$ denote the lowest eigenenergies from single-channel and coupled-channel calculations, respectively. $E_{th}^{Model}$ and $E_{th}^{Exp}$ are the theoretical and experimental thresholds. $E_{B}$ is the binding energy. "ub" indicates an unbound state.}
    \begin{tabular}{C{1.1cm} C{1.2cm}  C{1.8cm} C{1.6cm} C{1.5cm} C{1.8cm} C{1.6cm} C{1.5cm} C{2.0cm}}
            \toprule[1pt]
           $J^{P}$   & Channel  &  $E_{sc}$    &  $E_{th}^{Model}$   &$E_{B}$    &$E_{th}^{Exp}$    &$E_{sc}^{\prime}$   &$E_{cc}/E_{B}^{cc}$   &$E_{cc}^{\prime}$   \\
             \midrule[1pt]
             \multirow{5}{*}{$\frac{1}{2}^{-}$} &$\Lambda\eta_{s}$        &1979    &1975  &ub  &2074 &2078 &-55/1815 &1759 \\
                                                                       &$\Lambda\phi$               &2146    &2143  &ub  &2136 &2139 & \\
                                                                       &$\Xi K$                          &1867    &1869  &-2   &1813 &1811 &\\
                                                                       &$\Xi K^{\ast}$               &2261   &2266   &-5  &2210 &2205 &\\
                                                                       &$\Xi^{\ast} K^{\ast}$   &2392   &2388   &ub  &2425 &2429 &\\
             \midrule[1pt]
             \multirow{4}{*}{$\frac{3}{2}^{-}$} &$\Lambda\phi$                 &2147  &2143  &ub  &2136 &2140 &-28/1963 &2000\\
                                                                        &$\Xi K^{\ast}$               &2268  &2266  &ub  &2210 &2212 & &\\
                                                                       &$\Xi^{\ast} K$                &1990  &1991  & -1 &2028 & 2027& &\\
                                                                       &$\Xi^{\ast} K^{\ast}$    &2390  &2388  &ub  &2425 & 2427& &\\
           
            \midrule[1pt]
              $\frac{5}{2}^{-}$                           &$\Xi^{\ast} K^{\ast}$      &2370  &2388  &-18 &2425 &2407 &-18/2370 &2407\\
            \bottomrule[1pt]
        \end{tabular}
    \end{center}
\end{table*}

\begin{table*}[htb]
    \begin{center}
        \renewcommand{\arraystretch}{1.5}
        \caption{\label{bound-2} Calculated masses for the $I=1$ hidden-strange pentaquark system (in MeV). The notations are the same as in Table~\ref{bound-1}. }
     \begin{tabular}{C{1.1cm} C{1.2cm}  C{1.8cm} C{1.6cm} C{1.5cm} C{1.8cm} C{1.6cm} C{1.5cm} C{2.0cm}}
            \toprule[1pt]
            $J^{P}$   & Channel  &  $E_{sc}$    &  $E_{th}^{Model}$   &$E_{B}$    &$E_{th}^{Exp}$    &$E_{sc}^{\prime}$   &$E_{cc}/E_{B}^{cc}$   &$E_{cc}^{\prime}$   \\
            \midrule[1pt]
            \multirow{6}{*}{$\frac{1}{2}^{-}$} &$\Sigma\eta_{s}$            &2093    &2089  &ub  &2151 &2155 & ub/1873 &1816 \\
                                                                      &$\Sigma\phi$                   &2261    &2257  &ub  &2213 &2217 & \\
                                                                      &$\Sigma^{\ast}\phi$       &2382    &2379  &ub  &2405 &2408 & \\
                                                                      &$\Xi K$                           &1873    &1869  &ub   &1813 &1817 & \\
                                                                      &$\Xi K^{\ast}$                &2270   &2266   &ub  &2210 &2214 &\\
                                                                      &$\Xi^{\ast} K^{\ast}$    &2391   &2388   &ub  &2425  &2428 &\\
            \midrule[1pt]
            \multirow{6}{*}{$\frac{3}{2}^{-}$} &$\Sigma phi$                    &2261   &2257  &ub  &2213 &2216 &ub/1994 &2031\\
                                                                      &$\Sigma^{\ast}\eta_{s}$ &2215   &2211  &ub  &2343 &2347 &\\
                                                                      &$\Sigma^{\ast}\phi$        &2383   &2379  &ub  &2405 &2408 & \\
                                                                      &$\Xi K^{\ast}$                 &2270   &2266  &ub  &2210 &2214  & &\\
                                                                      &$\Xi^{\ast} K$                 &1995   &1991  &ub  &2028 &2032  & &\\
                                                                      &$\Xi^{\ast} K^{\ast}$     &2392   &2388  &ub  &2425 &2428  & &\\ 
            \midrule[1pt]
         \multirow{2}{*}{$\frac{5}{2}^{-}$}    &$\Sigma^{\ast}\phi$       &2383    &2379  &ub  &2405 &2409 &ub/2383 &2408  \\
                                                                     &$\Xi^{\ast} K^{\ast}$      &2392  &2388  &ub  &2425 &2429 &  \\
            \bottomrule[1pt]
        \end{tabular}
    \end{center}
\end{table*}

The preceding analysis of effective potentials has identified certain physical channels possessing effective attractive characteristics. Given this attraction, these channels are expected to form bound or resonance states. To verify the formation of such states within the dynamical estimation, we would subsequently perform a detailed estimation of their bound or resonance states properties.  Prior to presenting the dynamical results, it is essential to note the criteria for determining a bound state. In the single channel estimation, a bound state is identified if the lowest eigenvalue obtained is below the theoretical threshold of the corresponding channel. Similarly, in the multichannel coupling estimation, a bound state is indicated when the lowest eigenvalue is below the theoretical threshold of the lowest channel. It must also be emphasized that the dynamical estimation for the hidden-strange pentaquark system is performed within a finite space, which implies a limitation in the number of basis functions used. Consequently, we may obtain eigenvalues for some physical channels that involved repulsive properties. However, it is crucial to note that these are not bound state eigenvalues. Specifically, the lowest eigenvalue obtained would be higher than the theoretical threshold of the respective channels, and  these eigenvalues for repulsive channels would gradually converge towards the corresponding channel's theoretical threshold as the space is expanded. In contrast, a genuine bound state is independent of the system space expansion, and its stability is maintained throughout the strong decay process.

The estimated results are shown in Tables~\ref{bound-1} and ~\ref{bound-2}, which correspond to states with the $I=0$ and $I=1$, respectively.  In these tables,  $E_{sc}$ and $E_{cc}$ represent  the lowest eigenenergies obtained of the single channel and channel coupling estimation, respectively. $E^{model}_{th}$ and $E^{Exp}_{th}$ are the theoretical estimations and experimental measurements of the thresholds of the channels. Additionally, $E_{B}$ is the binding energy obtained by the formula of $E_{sc}-E_{th}^{model}$ and $E_{B}^{cc}$ stands for the binding energy obtained within the channel coupling estimations, which can be derived from the difference of the lowest eigenenergies obtained $E_{cc}$ in the channel coupling estimation and the theoretical threshold of the lowest channel in the system.  It should be noticed that the relevant parameters in the current work were determined by various aspects of hadron properties. Consequently, the uncertainty associated with these parameters would impact the accuracy of the model predictions. To reduce  model parameters dependence to certain extent, the calculated binding energy is utilized to correct both the single channel and channel coupling estimates. This procedure would yield the corrected single channel estimate $E_{sc}^{\prime}$ and the corrected channel coupling estimate $E_{cc}^{\prime}$, as presented in Table~\ref{bound-1} and Table~\ref{bound-2}, respectively. 

Table~\ref{bound-1} presents the estimated results of the dynamical calculation for the $I=0$ hidden-strange pentaquark system at various quantum numbers. Specifically, for the $J^{P}=1/2^{-}$ case, single channel estimation yields bound states for $\Xi K$ and $\Xi K^{\ast}$, with binding energies of approximately $-2$ MeV and $-5$ MeV, respectively. Consistent with the effective potential analysis in Fig.~\ref{I=0-effective}, the formation of these bound states is attributed to their attractive interaction. For the $\Xi^{\ast} K^{\ast}$ channel, although a strong attractive interaction exists between $\Xi^{\ast}$ and $K^{\ast}$, the lowest eigenenergy obtained from the bound state estimation is above its theoretical threshold, indicating that attraction is insufficient to form a bound state.  The $\Lambda \phi$ channel demonstrates a similar behavior, with its lowest eigenenergy also exceeding the corresponding theoretical threshold.  For the $\Lambda\eta_{s}$ channel, which is purely repulsive, the dynamical bound state estimation results in a lowest eigenenergy above its theoretical threshold. Consequently, the results for $\Xi^{\ast} K^{\ast}$, $\Lambda \phi$ and $\Lambda\eta_{s}$, having  eigenenergies above their threshold, are all identified as scattering states. In contrast, the multichannel coupling estimation reveals a deeply bound state with corrected mass of approximately 1758 MeV and a binding energy of approximately $-54$ MeV. This observation is consistent with the strongly attractive effective potential that emerges when channel coupling is considered, as shown in Fig.~\ref{I=0-effective}

For the $J^{P}=3/2^{-}$ case, the single channel estimation results shows that only channel $\Xi^{\ast}K$ becomes a shallow bound state with binding energy of around $-1$ MeV, driven by the attractive interaction within this channel. However, the single channel estimation for the remaining channels indicate their lowest eigenenergies are all higher than their corresponding theoretical threshold, signifying that these channels are identified as scattering states. Furthermore, a bound state with a binding energy of roughly $-28$ MeV and  a corrected mass of 2000 MeV is observed in  the channel coupling estimation,  The emergence of this bound state is due to the deep attractive effective potential within the channel coupling, as illustrated in Fig.~\ref{I=0-effective}.

For the $J^{P}=5/2^{-}$ case, since only the $\Xi^{\ast} K^{\ast}$ channel exists, the dynamical calculation shows that its attractive nature causes the lowest eigenenergy obtained for this channel to be below its theoretical threshold by about 18 MeV. This confirms the existence of a bound state in the $J^{P}=5/2^{-}$ system.

Table~\ref{bound-2} summarized the dynamical calculation results for the $I=1$ hidden-strange pentaquark system across different quantum numbers. In contrast to the $I=0$ system, the lowest eigenenergy obtained from all single channel estimation for the $I=1$ case, including $J^{P}=1/2^{-}$, $3/2^{-}$, and $5/2^{-}$, exceeds their respective theoretical threshold. Consequentially, as shown in Table~\ref{bound-2}, no bound states can be derived for all $I=1$ single channels.  This finding is in full agreement with the effective potential analysis presented in Fig.~\ref{I=1-effective}. In that analysis, the majority of $I=1$ channels display purely repulsive characteristic, or, where weak attraction is observed at the medium range, it is overwhelmed by a strong repulsive interaction at close range.  A bound state is similarly not found in the multichannel coupling estimation. For all quantum numbers, the lowest eigenenergies calculated from channel coupling remain above the theoretical threshold of the lowest channel. As a result, the consistency with Fig.~\ref{I=1-effective} further confirms that channel coupling within the $I=1$ system is either repulsive or its weak attractive interaction is inadequate to overcome the short-range repulsion, thereby precluding the formation of a bound state. 

 \begin{figure*}[t]
    \centering
    \includegraphics[scale=1]{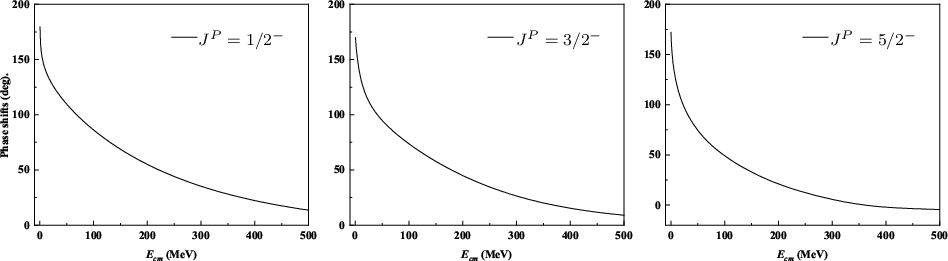}
    \caption{ The low energy scattering phase shifts for the $I=0$ hidden-strange pentaquark system for various quantum numbers in the coupling channel estimation. $E_{cm}$ denotes the center-of-mass incident energy. }
    \label{low-energy-phase-shifts}
\end{figure*}

\begin{table}[htb]
    \begin{center}
        \renewcommand{\arraystretch}{1.5}
        \caption{\label{low-phaseshifts} The scattering length $a_{0}$, the effective range $r_{0}$ and the binding energy $E_{B}^{\prime}$ determined by the variation method. }
    \begin{tabular}{C{1.1cm} C{1.2cm}  C{1.5cm} C{1.5cm} C{1.5cm} }
            \toprule[1pt]
          $I(J^{P})$ &    channel&    $a_{0}$ (fm)&     $r_{0}$ (fm)&     $E_{B}^{\prime}$ (MeV)  \\
            \midrule[1pt]
           $0(1/2^{-})$&    $\Xi K$&                          1.33&      0.63&          -56.9\\
           $0(3/2^{-})$&    $\Xi^{\ast} K$&              1.69&      0.74&          -29.9\\
           $0(5/2^{-})$&    $\Xi^{\ast} K^{\ast}$&  2.19&      1.0&          -17.7\\
            \bottomrule[1pt]
        \end{tabular}
    \end{center}
\end{table}
The results from the bound state estimation indicates the formation of three bound states with various quantum numbers in the $I=0$ hidden-strange pentaquark system.  To validate the existence of these three states, the low-energy scattering phase shifts behavior of the corresponding coupling channels by the well-developed Kohn-Hulthen-Kato (KHK) variational method detailed in Ref.~\cite{Kamimura:1977okl} is investigated, as shown in Fig.~\ref{low-energy-phase-shifts}. In accordance with scattering theory, the phase shifts approach $180^{\circ}$ as the incident energy tends toward zero, a behavior that serves as a definitive signature for the presence of a bound state.
Fig.~\ref{low-energy-phase-shifts} clearly demonstrates this behavior across the $J^{P}=1/2^{-}$, $3/2^{-}$, and $5/2^{-}$ quantum numbers, which strongly confirms the existence of three bound states within the channel coupling estimation.

Furthermore, based on the low-energy scattering information in Fig.~\ref{low-energy-phase-shifts}, we extracted key physical parameters, such as the scattering length $a_{0}$, effective range $r_{0}$, and binding energy $E_{B}^{\prime}$, using the formula (\ref{wave1}):
\begin{eqnarray}\label{wave1}
    k \cot{\delta_L} &=& -\frac{1}{a_{0}}+\frac{1}{2}r_{0}k^{2}+O(k^4),
\end{eqnarray}
where $k=\sqrt{2\mu E_{c.m}}$,  $\mu$ and $E_{c.m}$ correspond to the reduced mass of two hadrons and the incident energy, respectively. $\delta_{L}$ is the low-energy phase shifts obtained above. 

Subsequently, the results derived from the formula (\ref{wave1}) are substituted into the relation (\ref{wave2}) to determine the wave number $\alpha$, the formula \cite{Babenko:2003js} can be written by

\begin{eqnarray}\label{wave2}
    r_{0}&=&\frac{2}{\alpha}\left(1-\frac{1}{\alpha a_{0}} \right).
\end{eqnarray}

Finally, by inserting the calculated the wave number $\alpha$ into formula (\ref{wave3}), the binding energies are obtained by 
\begin{eqnarray}\label{wave3}
    E_{B}^{\prime}=\frac{\hbar^{2}\alpha^{2}}{2 \mu}.
\end{eqnarray}

The parameters determined using this approach are summarized in Table~\ref{low-phaseshifts}. The table clearly demonstrates that the scattering lengths 
at different quantum numbers are uniformly positive, and the resulting binding energies are consistent with the results from the dynamical bound estimation. This consistency further confirms the existence of three bound states in the $I=0$ hidden-strange pentaquark system. 
\begin{table*}[htb]
    \begin{center}
        \renewcommand{\arraystretch}{1.5}
        \caption{\label{each-interaction}Contributions of each terms in Hamiltonian to three bound states obtained with $I=0$ hidden-strange pentaquark system in unit of MeV. "SC" and "CC" represents the lowest single channel estimation and the channel coupling estimation, respectively.  }
            \begin{tabular}{C{2.0cm} C{2.0cm}  C{2.0cm} C{2.0cm} C{2.0cm} C{2.0cm} C{2.0cm} C{2.0cm}}
            \toprule[1pt]
            $J^{P}$&          SC/CC&       $H_{T}$&       $V_{CON}$&      $V_{OGE}$&    $V_{\pi}$&   $V_{\eta}$&  $V_{K}$ \\ 
            \midrule[1pt]
            \multirow{2}{*}{$\frac{1}{2}^{-}$}&     SC&          14.3&    -8.6&    -9.7&      2.1&    -0.5&   0.4 \\ 
                                                                     &     CC&        100.7&   -86.5&  -60.4&  -16.3&     5.1&    2.7  \\
            \midrule[1pt]
            \multirow{2}{*}{$\frac{3}{2}^{-}$}&     SC&          8.9&    -7.5&    -4.3&     1.4&        -0.06&   0.5   \\
                                                                     &     CC&          75.7&   -78.2&  -21.7&  -11.6&      0.2&    7.2     \\
            \midrule[1pt]                                                       
              $\frac{5}{2}^{-}$                          &      SC&     -7.9&  -13.7&   -4.1&    7.7&         -0.8&     1.2  \\
            \bottomrule[1pt]
        \end{tabular}
    \end{center}
\end{table*}

The preceding analysis conclusively demonstrates that for the $I=0$ hidden-strange pentaquark system with different quantum number the lowest eigenenergies obtained 
in the single channel and channel coupling estimation are below the corresponding lowest threshold channel. To find which interaction plays the dominant role in forming a below-threshold bound state and further gain a more quantitative understanding of the 
influence of channel coupling on individual interaction terms, we list the contribution of each term for both the lowest threshold single channel and the channel coupling corresponding to the three determined bound states. The corresponding results are presented in Table~\ref{each-interaction}.

From the Table~\ref{each-interaction},  it is noted that the formation of bound states arises from the interplay of multiple interaction.  For the case of $J^{P}=1/2^{-}$ and $3/2^{-}$ system,  although the short-range $V_{CON}$ and $V_{OGE}$ provide fundamental attraction in the single channel estimation, they are insufficient to form deeply bound states. The channel coupling effect plays a decisive role here. Quantitatively, it significantly enhances the magnitude of the short-range attraction from $V_{CON}$ and $V_{OGE}$. Qualitatively, it activates a critical attractive contribution from pion exchange ($V_{\pi}$), transforming it from a repulsive or negligible term in the single-channel case into a significant attractive factor. It is this superposition of the enhanced short-range forces and the induced meson-exchange attraction that ultimately overcomes the repulsive terms, leading to the formation of deeply bound states. For the case of $J^{P}=5/2^{-}$, which involves only a single physical channel, the formation of the bound state is primarily driven by attractive contributions from the confinement potential $V_{CON}$, the one-gluon exchange potential $V_{OGE}$, and the kinetic energy term $H_{T}$, as shown in Table~\ref{each-interaction}. The attractive nature of the kinetic energy is a characteristic feature of the QDCSM, attributed to the quark delocalization effect which effectively lowers the kinetic energy of the system.



\subsection{Scattering process and resonance states}
 \begin{figure*}[t]
    \centering
    \includegraphics[scale=0.25]{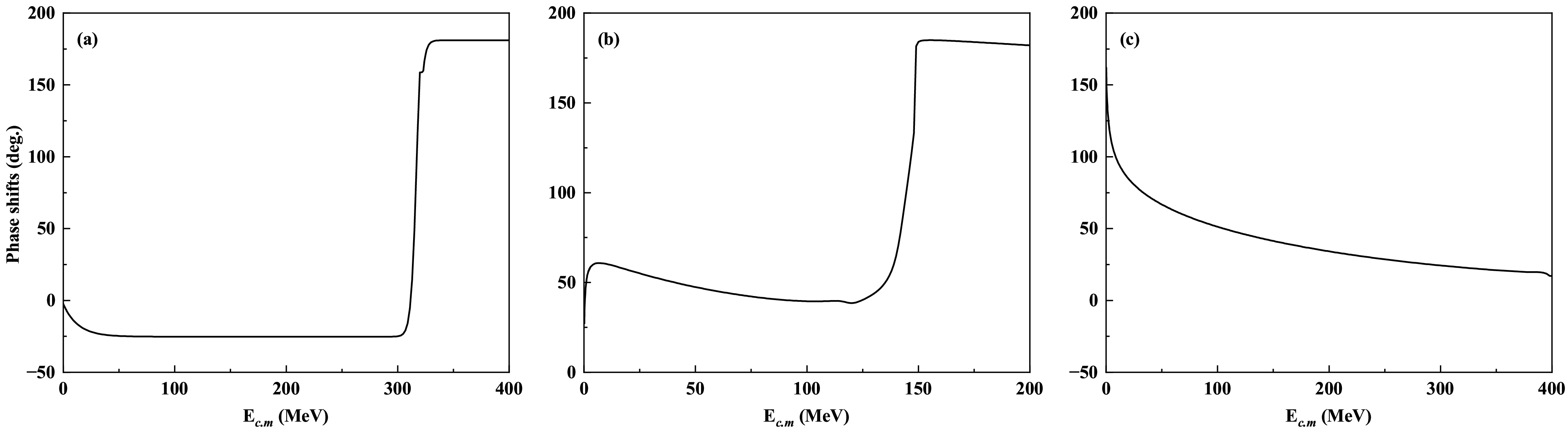}
    \caption{The phase shifts of the open channels with two-channel coupling for I=0 in QDCSM. (a) corresponds to two-channel coupling with $\Lambda \eta_{s}$ and $\Xi K^{\ast}$, (b) denotes two-channel coupling with $\Lambda\phi$ and $\Xi K^{\ast}$, (c) is the two-channel coupling with $\Xi K$ and $\Xi K^{\ast}$.  }
    \label{phase-shifts}
\end{figure*}

\begin{table}[htb]
    \begin{center}
        \renewcommand{\arraystretch}{1.5}
        \caption{\label{decaywidth}The mass and decay widths (in the unit of MeV) of resonance states $\Xi K^{\ast}$ with difference scattering process for $I=0$ hidden-strange pentaquark system. $M^{th}_{R}$ represents the sum of the corresponding theoretical threshold of the open channel and the incident energy. $M^{co}_{R}$ stands for the modified resonance mass. $\Gamma_{cc}$ is the width of the resonance state decaying into the certain open channel. $\Gamma_{hd}$ is the widths caused by the decay of the hadron components in the resonance states. $\Gamma_{total}$ can be obtained by the relation $\Gamma_{total} = \Gamma_{cc} + \Gamma_{hd}$ }
     \begin{tabular}{C{2.5cm} C{1.cm}  C{1.cm} C{1.cm} C{1.cm} C{1.cm} }
            \toprule[1pt]
          open channels&   $M^{th}_{R}$&    $M^{co}_{R}$&     $\Gamma_{cc}$&   $\Gamma_{hd}$  & $\Gamma_{total}$\\
                      \midrule[1pt]
            $\Lambda\eta_{s}$&     2265&     2208&      5.5&      50.2&   55.7 \\
            $\Lambda \phi$      &     2260&     2204&    12.4&      50.9&   63.3 \\                                 
            \bottomrule[1pt]
        \end{tabular}
    \end{center}
\end{table}
Given the above effective potential and bound state estimation results analysis, several bound states in the single channel estimation have been identified. These bound states can couple to specific open channels, through which they may manifest as resonance or scattering states in the corresponding open channel phase shifts. Typically, a sharp rise observed in the scattering phase shifts of an open channel coupled to a bound state indicates that the bound state has transformed into a resonance. Conversely, the absence of such behavior suggests the bound state 
may have dissolved into the continuum or become a scattering state. Additionally, since no bound states are found in the $I=1$ hidden-strange pentaquark system in single channel and coupling channel estimations, our investigation is restricted to the $I=0$ hidden-strange pentaquark system. To investigate the existence of resonance states in this system, all possible open channels scattering processes are analyzed. Fig.~\ref{phase-shifts} presents the scattering phase shifts behavior for all possible open channels. Table~\ref{decaywidth} summarizes the masses and decay widths of all potential resonance states. Notably, the current estimation neglects contributions from higher waves, considering only the decay widths into $S-$wave open channels. 

For the $J^{P}=1/2^{-}$ case, two bound states, $\Xi K$ and $\Xi K^{\ast}$, are identified in the single channel estimation. The lowest eigenenergy of bound state $\Xi K^{\ast}$ lies above the threshold of the lowest open channel. This implies that $\Xi K^{\ast}$ can couple to channels with threshold below its energy.  It is therefore necessary to investigate whether bound state $\Xi K^{\ast}$ manifests as a resonance in the scattering phase shifts of these specific open channels due to the channel coupling effect. For the bound state $\Xi K$, however, its lowest eigenenergy is below the lowest channel threshold of the system. Given that only $S-$wave are considered in the current framework, no corresponding open channel are energetically accessible for this state.

The bound state $\Xi K^{\ast}$ can decay into open channels with lower thresholds, such as $\Lambda\phi$, $\Lambda\eta_{s}$ and $\Xi K$. As shown in Fig.~\ref{phase-shifts}, a resonance signal corresponding to the $\Xi K^{\ast}$ state  is observed in the scattering phase shifts of open channels $\Lambda\eta_{s}$  and $\Lambda\phi$; however, no such signal is found in $\Xi K$ channel. This suggests that the  $\Xi K^{\ast}$ bound state manifests as a resonance primarily in scattering phase shifts of open channels $\Lambda\eta_{s}$  and $\Lambda\phi$.  Furthermore, an analysis of the scattering phase shifts for channels $\Xi K$ reveals that the phase shifts approach $\pi$ as the incident energy approaches zero. This behavior indicates the existence of a new bound state near the threshold of channel $\Xi K$, induced by the channel coupling effect.

By analyzing the scattering phase shifts,  the mass and the decay width of the identified resonance state are obtained, as presented in Table~\ref{decaywidth}.  According to Table~\ref{decaywidth}, the resonance state mass $M_{R}^{th}$ is defined as the sum of the open channel threshold and the incident energy at which the phase shift reaches $\pi/2$, while $M^{co}_{R}$ indicates the corrected resonance mass, calculated through $M^{co}_{R}=M_{R}^{th}-\sum_{i}p_{i}(E_{th}^{Model}(i)-E_{th}^{Exp}(i))$, where $p_{i}$ denotes the contribution of the $i$th channel in the channel coupling estimation. The scattering decay width is determined by the difference in incident energies corresponding to phase shifts of $\pi/4$ and $3\pi/4$.  Notably, the intrinsic decay widths of the constituent baryons or mesons, particularly for hadrons with large widths, significantly impact the resulting resonance parameters.  To incorporate this effect, the decays of these constituent hadrons are also considered. Therefore, the total decay width $\Gamma_{total}$ is estimated using the relation $\Gamma_{total} = \Gamma_{cc} + \Gamma_{hd}$, where $\Gamma_{cc}$ is the decay width obtained from the coupled-channel scattering process, and $\Gamma_{hd}$ is the width arising from the intrinsic decays of the constituent baryons or mesons within the pentaquark state. $\Gamma_{hd}$ can be calculated following the method provided in Ref.~\cite{Ping:2008tp}.   

For resonance state $\Xi K^{\ast}$,  the intrinsic decay width  $\Gamma_{hd}$ is dominated by the decay of the constituent $K^{\ast}$ meson.  We estimate this width by scaling the free-space width ($\Gamma_{fK^{\ast}} \approx 51$ MeV) to account for the reduction in phase space available to the bound $K^{\ast}$. The effective mass of the bound $K^{\ast}$ is reduced approximately to:
\begin{eqnarray}
  M_{bK^{\ast}}\approx M_{fK^{\ast}}-\frac{m_{K^{\ast}}}{m_{\Xi}+m_{K^{\ast}}}E_{B}
\end{eqnarray}
where $E_{B}$ is the binding energy of the bound state. The modified decay width is then given by: 
\begin{eqnarray}
   \Gamma_{bK^{\ast}}(M_{bK^{\ast}}) \approx \Gamma_{fK^{\ast}}\frac{k^{2l}_{b}\rho(M_{bK^{\ast}})}{k^{2l}_{f}\rho(M_{fK^{\ast}})}
\end{eqnarray}
where $k$ is the momentum of the decay pion in the $K^{\ast}$ rest frame, and $l=1$ represents the orbital angular momentum of the $P$-wave decay. The two-body phase space factor $\rho(M)$ is defined as:
\begin{eqnarray}
    \rho(M)=\pi \frac{kE_{\pi}E_{K}}{M}
\end{eqnarray}
where $E_{i}$ ( i=$\pi, K$) is the center-of-mass energy of the corresponding decay product. Using these formulas, the width $\Gamma_{hd}(\Xi K^{\ast})$ (approximated as $\Gamma_{b K^{\ast}}$) is calculated to be in the range of $48\text{--}50$ MeV. The uncertainty in the binding energy, derived from different open-channel scattering processes, propagates to the decay width estimation. Consequently, the corrected mass of the $\Xi K^{\ast}$ resonance is predicted to be  $2204\text{--}2208$ MeV, with a total decay width of approximately $55.7\text{--}63.3$ MeV.. 

For the $J^{P}=3/2^{-}$ case, a bound state $\Xi^{\ast} K$ is derived in the single-channel estimation, and its eigenenergy lies below the lowest physical channel threshold of system. This implies that within the current framework which only considers $S$-wave, no energetically allowed decay channels are available for this state. Consequently, it appears as a stable bound state rather than a resonance in this framework. Similarly, the $J^{P}=5/2^{-}$ $\Xi^{\ast} K^{\ast}$ state is found to be deeply bound below the lowest threshold. 
While strong decay via $D$-wave transitions is physically possible, such processes are beyond the scope of the current $S$-wave study. Therefore, these states are predicted to be stable against $S$-wave strong decays. 


\section{Summary\label{sum}}

Recently, the BESIII Collaboration reported the first study of the decay process $\psi(3686)\rightarrow \Sigma^{0}\bar{\Sigma}^{0}\phi$ with a statistical significance of $7.6 \sigma$, and searched for intermediate resonance states in the $\Sigma^{0}\phi(\bar{\Sigma}^{0}\phi)$ invariant mass distribution~\cite{BESIII:2024muk}. Although no clear signal was observed due to limited statistics,  the existence of such states remains an open question. Motivated by this measurement, we perform a systematic study of the hidden-strange pentaquark system using the resonating group method within the QDCSM framework.  

In the present work, we first investigate the effective potential of each channel across different quantum numbers to determine their interaction properties. Subsequently, both single-channel and multichannel coupling estimations are performed to search for possible stable bound states.  The single channel estimation presents two bound states in the $I(J^{P})=0(1/2^{-})$ sector: a  $\Xi K$ with an estimated mass of 1811 MeV and a $\Xi K^{\ast}$ with an estimated mass of 2205 MeV. Additionally, bound states are found in the $\Xi^{\ast} K$ channel ($I(J^{P})=0(3/2^{-})$) and the $\Xi^{\ast} K^{\ast}$ channel ($I(J^{P})=0(5/2^{-})$),  with estimated masses of 2027 MeV and 2407 MeV, respectively.  After accounting for the effects of multichannel coupling,  three hidden-strange pentaquark states with $I(J^{P})$ quantum numbers of $0(1/2^{-})$, $0(3/2^{-})$, and $0(5/2^{-})$ are predicted, with estimated masses of approximately $1759$ MeV, $2000$ MeV, and $2407$ MeV, respectively. 

Furthermore, based on the bound state estimation, the scattering phase shifts of various open channels are analyzed to search for possible resonance states. The analysis reveals the existence of a resonance state dominated by the $\Xi K^{\ast}$ channel in the $I(J^{P})=0(1/2^{-})$ sector. This resonance state manifests clearly in the scattering phase shifts of the $\Lambda\eta_{s}$ and $\Lambda\phi$ channels.  Considering both the scattering width from channel coupling and the intrinsic decay width of the constituent $K^{\ast}$ meson, the mass and total decay width of this resonance are predicted to be $2204\text{--}2208$ MeV and $55.7\text{--}63.3$ MeV, respectively. These theoretical predictions provide specific targets and important guidance for future experimental searches for hidden-strange pentaquarks at facilities such as BESIII.

\acknowledgments{This work is supported partly by the National Natural Science Foundation of China under
    Contract No. 12175037, No. 12335001, No. 11775118 and No. 11535005, and is also supported by the Natural Science Foundation for Youths of Henan Province  No. 252300421781.  School-Level Research Projects of Henan Normal university (No. 20240304) also supported this work. Besides, Jiangsu Provincial Natural Science Foundation Project (No. BK20221166), National Youth Fund: No. 12205125 and School-Level Research Projects of Yancheng Institute of Technology (No. xjr2022039, xjr2025010) also supported this work.}

\bibliography{uss-ssbar}

\end{document}